\documentclass[a4paper,nobalancelastpage,twocolumn,
superscriptaddress,prl,amsmath,amssymb,showpacs]{revtex4}

\usepackage{graphicx}

\newcommand{\ket}[1]{|#1\rangle}
\newcommand{\bra}[1]{\langle#1|}
\newcommand{\tr}{\mathrm{tr}}

\begin{document}

\title{Entropy scaling and simulability by Matrix Product States}

\author{Norbert Schuch}
\affiliation{Max-Planck-Institut f\"ur Quantenoptik,
  Hans-Kopfermann-Str.\ 1, D-85748 Garching, Germany.}
\author{Michael M.\ Wolf}
\affiliation{Max-Planck-Institut f\"ur Quantenoptik,
  Hans-Kopfermann-Str.\ 1, D-85748 Garching, Germany.}
\author{Frank Verstraete}
\affiliation{Fakult\"at f\"ur Physik, Universit\"at Wien, 
Boltzmanngasse 5, A-1090 Wien, Austria.} 
\author{J.\ Ignacio Cirac}
\affiliation{Max-Planck-Institut f\"ur Quantenoptik,
  Hans-Kopfermann-Str.\ 1, D-85748 Garching, Germany.}


\begin{abstract}
We investigate the relation between the scaling of block entropies and the
efficient simulability by Matrix Product States (MPS), and 
clarify the connection both for von Neumann and R\'enyi entropies, as
summarized in Table~\ref{table:intensive}. Most notably, even states
obeying a strict area law for the von Neumann entropy are not
necessarily approximable by MPS.  We apply these results to illustrate that 
quantum computers might outperform classical computers in simulating the
time evolution of quantum systems, even for completely translational
invariant systems subject to a time independent Hamiltonian. 
\end{abstract}

\maketitle

Understanding the behaviour of quantum many-body systems is a central
problem in physics.  Recently, Matrix Product States (MPS) have received
much interest as a variational ansatz for the simulation of
correlated one-dimensional systems.  They have proven particularly
powerful in approximating the  ground states of local Hamiltonians, as
used in the DMRG method~\cite{white:DMRG-PRL,schollwoeck:rmp}, but have
also been applied, e.g., to simulate the time evolution of slightly
entangled quantum systems~\cite{vidal:timeevol}.  Despite considerable
progress~\cite{hastings}, it is still not fully understood which
property exactly a state has to fulfil to be well approximated by MPS.  This
knowledge is not only of practical interest, but could also tell
us how to extend the MPS ansatz to, e.g., higher dimensional systems.

\begin{table}[b]
\includegraphics[width=8cm]{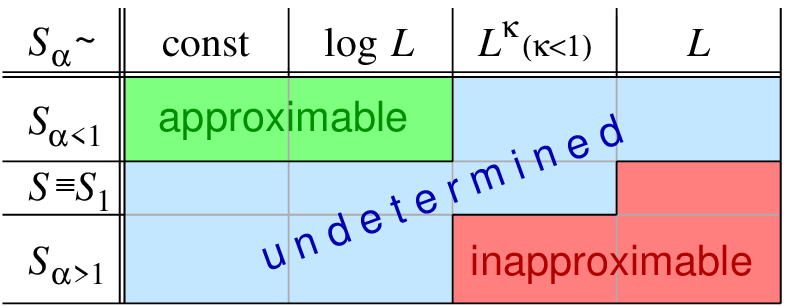}
\caption{
\label{table:intensive}
Relation between scaling of block R\'enyi entropies and approximability by
MPS.  In the ``undetermined'' region, 
nothing can be said about approximability just from looking at the
scaling.
}
\end{table}

It is generally believed that the relevant criterion for efficient
approximability by MPS is that the states under consideration obey an area
law, i.e., the von Neumann entropy of a block is bounded.   Although
indeed both ground states of local Hamiltonians and MPS obey an area law,
there are reasons to doubt this immediate connection: Firstly, the von
Neumann entropy is an asymptotic concept, quantifying what happens when
dealing with a large number of copies of a state.  Conversely, it
has been shown recently that a rigorous connection can be established by
looking at R\'enyi entropies instead~\cite{faithfully}. Unfortunately, the
argument used breaks down as the von Neumann entropy is approached.
Finally, the continuity inequality for the von Neumann entropy carries a
size-dependent constant, and thus states which are close to each other 
need not be close in entropy~\cite{audenaert:fannes}.

In this work, we explore the connection between entropy scaling and
approximability by MPS.  The results are summarized in
Table~\ref{table:intensive}: An at most logarithmic scaling of R\'enyi
entropies $S_\alpha$, $\alpha<1$, implies approximability by MPS.  On
the other side, a faster than logarithmic increase of $S_\alpha$,
$\alpha>1$, rules out efficient approximability by MPS, as does linear
growth of the von Neumann entropy.  For all other cases, the scaling of
the block entropy does not allow for conclusions about approximability.
In particular, this holds for the case of constant von Neumann entropy,
which demonstates that the reason why MPS describe ground states well is
not simply that those states obey an area law.

Finally, we apply our results to illustrate that quantum computers might
outperform classical computers in simulating time evolutions. It is
long-known that quantum computers can simulate the behavior of quantum
systems~\cite{lloyd}. However, this does not automatically imply that they
will outperform classical computers, as, e.g., ground states of gapped
quantum systems appear classically efficiently
approximable~\cite{hastings,hastings2}.  On the other hand, it is known
that time evolution even of one-dimensional systems under a translational
invariant Hamiltonian can implement quantum computations if either
translational invariance is broken by the initial, boundary, or final
conditions, or the Hamiltonian is time dependent~\cite{bqp-evol}, and is
thus hard to simulate.  We extend these results by showing that even the
simplest case, the evolution of a translational invariant spin $\tfrac12$
system with translational invariant initial conditions under a time
independent Hamiltonian, cannot be simulated efficiently using MPS; this
provides evidence that quantum computers might outperform classical
computers in simulating these systems.

Let us first introduce the relevant quantities and notations.  We want to
obtain approximations which reproduce accurately not only the local
properties such as energy, but also the non-local ones such as
correlations.  This is ensured by bounding the error made
between two states $\ket{\psi}$ and $\ket\phi$
for an arbitrary observable $O$,
\[
\left|\tr[\psi O]-\tr[\phi O]\right|
    \le\|O\|_\mathrm{op}\|\psi-\phi\|_\tr\ ,
\]
where throughout the paper, $\psi\equiv\ket\psi\bra\psi$ etc.\ 
denotes the corresponding density
operator. We focus on non-extensive
observables (see Footnote~\footnote{
For extensive observables, where $\|\psi-\psi_D\|_\tr\le\delta/N$,
the results are the same except that $S\sim N^\kappa$ now 
implies inapproximability [replace $\delta$ by $\delta/N$ in
(\ref{eq:thm2:bondbound})]. In the approximability example for linearly
growing $\alpha<1$ R\'enyi entropy, one has to set $p_N=1/N^3$. 
As an example of an inapproximable state with bounded von Neumann
entropy, a single copy of (\ref{eq:magicstate}) is now sufficient, 
yielding a translation invariant example.
} for extensive observables),  
therefore w.l.o.g.\ $\|O\|_\mathrm{op}\le1$.
It follows that by imposing
\begin{equation}
\label{eq:before-trig}
\|\psi-\phi\|_\mathrm{tr}\le\delta\ ,
\end{equation}
we bound the error made in any observable by $\delta$.

For some of the proofs it will be more convenient to consider the two-norm
distance $\big\|\ket\psi-\ket\phi\big\|_2$ or the fidelity 
$\big|\langle\phi\ket\psi\big|/\big\|\ket\phi\big\|_2
\big\|\ket\psi\big\|_2=:\cos(\theta)$. Fortunately, these measures turn all
out to be equivalent: Since the best approximating MPS will generally not
be normalized, it is appropriate to consider the optimized quantities, and
one finds that 
$
T(\phi,\psi):=
\inf_{\alpha}\|\psi-\alpha\phi\|_\tr/\|\psi\|_\tr\equiv 
\sin(2\theta)$  and 
$V(\phi,\psi):=\inf_\alpha
\big\|\ket{\psi}-\alpha\ket\phi\big\|_2/\|\ket{\psi}\|_2
\equiv\sin(\theta)$ for $0\le\theta\le\tfrac\pi4$.

We now introduce Matrix Product States (MPS)~%
\cite{ksz,vidal:timeevol,mps-reps}.
Consider a chain of
$N$ $d$-level systems with the corresponding Hilbert space 
\[\mathcal H_N:=(\mathbb C^d)^{\otimes N}\ .\]
We call $\ket{\phi_D} \in\mathcal H_N$ a
Matrix Product State (MPS) with \emph{bond dimension} $D$ (or, briefly, a
$D$-MPS) if it can be written as
\begin{equation}
\label{eq:mps-rep}
\ket{\phi_D}=\sum_{i_1,\ldots,i_N=1}^d A^{[1]}_{i_1}
A^{[2]}_{i_2}\cdots A^{[N]}_{i_N}\ket{i_1,i_2,\ldots,i_N}
\end{equation}
with 
$A^{[k]}_i$ $D\times D$ matrices for $2\le k\le N-1$, and $A^{[1]}_i$ and
$A^{[N]}_i$ row and column vectors of length $D$,
respectively~\footnote{Using MPS with periodic boundary conditions (PBC)
gives the same results, as any PBC MPS with bond dimension $D$ can be
embedded in an MPS with open boundaries (\ref{eq:mps-rep}) with bond
dimension $D^2$.}.

Given a family
$(\ket{\psi_N})\equiv(\ket{\psi_N})_{N\in\mathcal N\subset\mathbb N}$ of states,
$\ket{\psi_N}\in\mathcal H_N$, we say that it \emph{can be 
approximated efficiently by
MPS} if for every $\delta>0$, there exists a sequence 
${\ket{\phi_{N,D}}}$ of MPS 
with $D\equiv D(N)=O(\mathrm{poly}_\delta(N))$
such that $\|{\psi_N}-{\phi_{N,D}}\|_\tr\le\delta$.
 On the contrary, if there is some
$\delta>0$ such that no sequence of MPS with polynomial bond dimension
can approximate $\ket{\psi}$ up to $\delta$, we say that
$(\ket{\psi_N})$ \emph{cannot be approximated efficiently by
MPS}. For brevity, we will sometimes drop the word ``efficiently''.

 We will measure entropies using the \emph{R\'enyi entropies}
\[
S_\alpha(\rho)=\frac{\log \tr \rho^\alpha}{1-\alpha}\ ,\ 0\le
    \alpha\le\infty\ ,
\]
which are a generalization of the von Neumann entropy
$S(\rho)=-\tr[\rho\log\rho]$.  In particular,
$\lim_{\alpha\rightarrow1}S_\alpha(\rho)=S(\rho)$.
Note that all logs are to the basis $2$.

We aim to relate approximability by MPS to the scaling of block
entropies. To this end, we first show that the error made in approximating
some state by a $D$-MPS is determined by the error made when truncating
the Schmidt spectrum of its bipartitions after $D$ values.  Therefore, let
$\ket\psi\in\mathcal H_N$, $\rho_k=\tr_{k+1,\dots,N}\ket{\psi}\bra{\psi}$,
and let $\lambda^{[k]}_1\ge\lambda^{[k]}_2\ge\cdots\ge\lambda^{[k]}_{d^k}$
be the ordered spectrum of $\rho_k$. Then, define the \emph{truncation
error} $$\epsilon_k(D):=\sum_{i=D+1}^{k^d}\lambda^{[k]}_i\ .$$

Let us now relate the truncation error to approximability
by MPS.  The intuition is that the best an MPS with bond dimension $D$
(i.e., Schmidt rank $D$ in any bipartition) can do is to preserve the
$D$ largest eigenvalues, resulting in an error of $\epsilon_k(D)$
for the cut at $k$ (which can, but need not, accumulate). On the one side,
it has been shown in \cite{faithfully} that for a state
$\ket\psi\in\mathcal H_N$, there always exists an MPS $\ket{\phi_D}$ with
bond dimension $D$ such that
\begin{equation}
\label{eq:eps-upper}
\big\|\ket\psi-\ket{\phi_D}\big\|_2\le2\sum_{k=1}^{N-1}\epsilon_k(D)\ .
\end{equation}
On the other hand, any $D$-MPS $\ket{\phi_D}$ 
satisfies 
\begin{equation}
\label{eq:eps-lower}
\|\psi-\phi_D\|_\tr\ge\epsilon_k(D)\quad\forall k\ ,
\end{equation}
since 
with
$\rho_{k}=\tr_{1,\dots,k}\psi$ and
$\sigma_{D,k}=\tr_{1,\dots,k}\phi_D$, 
\[
\|\psi-\phi_D\|_\tr\ge 
\|\rho_k-\sigma_{D,k}\|_\tr
\ge\epsilon_k(D)\ .
\]
Here, we have used: \textit{i)} the contractivity of the partial trace,
\textit{ii)} for fixed spectra, the trace norm distance is extremal for
commuting operators~\cite{bhatia}, and \textit{iii)}
$\mathrm{rank}\,\sigma_{D,k}\le D$.

We start the discussion of Table~\ref{table:intensive} by proving the
cases for which conclusive statements can be made.  In the following,
$\rho_N^L$ will denote any $L$-particle reduced
block of a state $\ket{\psi_N}\in\mathcal H_N$.
The case of at most logarithmically growing R\'enyi entropy with
$\alpha<1$ was discussed in \cite{faithfully}, where is was shown that it
implies approximability.  More formally, if for a family of states
$(\ket{\psi_N})$ there exist $c,c'>0$ and $0\le\alpha<1$ such that
$S_\alpha(\rho_N^L)\le c \log(N)+c'$ for all reduced blocks $\rho_N^L$,
then it can be approximated efficiently by MPS.

Let us now show that a linearly growing von Neumann entropy implies
inapproximability. Formally, 
if for a family $(\ket{\psi_N})$, $S(\rho_N^L)\ge c L$ holds for some
$c>0$, $L\equiv L(N)\ge\eta N$, $\eta>0$, and some reduced blocks
$\rho_N^L$, then it cannot be approximated
efficiently by MPS.

To prove this, we use Fannes' inequality in its improved version by
Audenaert~\cite{audenaert:fannes}: For density operators $\rho$, $\sigma$ on a
$K$-dimensional Hilbert space, 
$
|S(\rho)-S(\sigma)|\le T\log (K-1)+H(T,1-T)
$,
where $2T=\|\rho-\sigma\|_\tr\le\delta$, and $H(T,1-T)\le1$ is the binary
entropy.  Let $(\ket{\phi_{N,D}})$ be a sequence of MPS 
approximating $(\ket{\psi_N})$, and $\rho_N^L$, $\sigma_{N,D}^L$ the
corresponding reduces states for which $S(\rho_N^L)\ge c L$. Then,
\[
|S(\rho_N^L)-S(\sigma_{N,D}^L)|\le\tfrac12 \delta L\log d+1\ ,
\]
and thus, for $L\ge\eta N$,
\begin{align}
\label{eq:thm2:bondbound}
\log D(N)\ge S(\sigma_{N,D}^L)&\ge S(\rho_N^L)-\tfrac12  \delta L \log d -1\\ 
&\ge \eta(c-\tfrac12\delta\log d)N-1\ ,\nonumber
\end{align}
i.e., the bond dimension grows exponentially in $N$ as soon as
the error $\delta<2c/\log d$, which completes the proof.

In the following, we show that a faster than logarithmic increase of any
R\'enyi entropy with $\alpha>1$ also implies inapproximability, i.e.,  
if for a family $(\ket{\psi_N})$, there exist
$\alpha>1$ and $\kappa>0$ s.th.\ $S_\alpha(\rho_N^L)\ge c
L^\kappa$ for some $c>0$, $L\equiv L(N)\ge \eta N$, and
some reduced blocks $\rho_N^L$, then
it cannot be approximated efficiently by MPS.

This is proven by lower bounding the truncation error
$\epsilon\equiv\epsilon(D)$ of a block $\rho^L_N$ for given
$S_\alpha(\rho)$ ($\alpha>1$) and then applying
(\ref{eq:eps-lower}).  This, however, is the same as maximizing
the entropy while keeping $\epsilon$ fixed. Since both the entropy and
$\epsilon$ only depend on the spectrum, the problem reduces to a
classical one. It is easy to see that the probability distribution
$$
p_1,\ldots,p_D=\frac{1-\epsilon}{D};\ p_{D+1},\ldots,p_{2^L}=
    \frac{\epsilon}{2^L-D}
$$
is majorized by all ordered probability distributions $(q_i)$ which
satisfy $q_{D+1}+\dots+q_{2^N}=\epsilon$, and since R\'enyi entropies are
Schur concave functions, it has maximal entropy~\cite{bhatia}.
Therefore, we obtain the inequality
\begin{align*}
S_\alpha&(\rho_N^L)\le 
    \frac{-1}{\alpha-1}\log\left[\frac{(1-\epsilon)^\alpha}{D^{\alpha-1}}+
	    \frac{\epsilon^\alpha}{(2^L-D)^{\alpha-1}}\right]\\
    &\le
    \frac{-1}{\alpha-1}\log\left[
	    \frac{(1-\epsilon)^\alpha}{D^{\alpha-1}}\right]
    = \log D-\frac{\alpha}{\alpha-1}\log(1-\epsilon)\;.
\end{align*}
Since from (\ref{eq:eps-lower}) the total error is
$\delta\ge\epsilon$, we find
\[
\log D\ge
S_\alpha(\rho_N^L)+\frac{\alpha}{\alpha-1}\Big|\log(1-\delta)\Big|\ ,
\]
and from $S_\alpha(\rho_N^L)\ge c L^\kappa\ge c\eta^\kappa
N^\kappa$, we infer that $D$ has to grow exponentially for any $\delta$.

We now turn towards the undetermined region in
Table~\ref{table:intensive}, where we provide examples for both
approximability and inapproximability. This task is greatly simplified by
the fact that approximability examples extend to the top and
left in Table~\ref{table:intensive}, while inapproximability extends to
the right and bottom. This holds as approximability for a given scaling
implies the same for more moderate scalings (and conversely for
inapproximability), and since $S_\alpha(\rho)$ decreases monotonically
in $\alpha$.  

The aim of this work is to clarify the relation between entropy scaling
laws and the approximability by MPS: Therefore, our examples are not
constructed to be ground states.  Yet, all of them form uniform families
of states, i.e., they can be generated by a uniform family of time
dependent Hamiltonians. The existence of time-independent realizations is
plausible, as the central ingredient of the examples are properly
distributed entangled pairs. These could be represented by pairs of
localized excitations which are prepared locally and then propagated by a
time-independent Hamiltonian.  

All of the examples can be chosen to be translational invariant, with the
only possible 
exception of the inapproximability example for constant von Neumann
entropy. The question whether any translational invariant state with
bounded von Neumann entropy can be approximated efficiently by MPS thus 
remains
open. 

The examples can be grouped into two classes; the first is based on
states of the type
\begin{equation}
\ket{\psi_{2N}}=\sqrt{1-p_{N}}\ket{2}^{\otimes 2N}+
    \sqrt{\frac{p_{N}}{2^{N}}}\sum_{x\in\{0,1\}^{N}}\ket{x}\ket{x}\ .
\label{eq:magicstate}
\end{equation}
By choosing $p_N=1/N$, we obtain an example of a state with linearly growing
R\'enyi entropies for all $\alpha<1$ which can be approximated by MPS,
as 
$$\big\|\ket{\psi_{2N}}-\sqrt{1-p_{N}}\ket{2}^{\otimes 2N}\big\|_2=
\sqrt{p_N}\rightarrow0\ .$$  
On the other hand, for $L\le N$,
\[
\rho_{2N}^L=(1-p_N)\ket{2}\bra{2}^{\otimes L}+
    \frac{p_N}{2^L}\sum_{y\in\{0,1\}^L}\ket{y}\bra{y}\ ,
\]
and therefore
\begin{align*}
S_\alpha(\rho_{2N}^L)&=\frac{1}{1-\alpha}
	\log\left[(1-p_N)^\alpha+2^{(1-\alpha)L}p_N^\alpha\right]\\
    &\ge L-\frac{\alpha}{1-\alpha}\log N\ .
\end{align*}
Note that the infavourable scaling of 
$c_\alpha:=\frac{\alpha}{1-\alpha}$ for $\alpha\rightarrow1$ can be
compensated by e.g.\ choosing $p_N=N^{-1/c_\alpha}$.

The next example provides states with algebraically (but sublinearly)
growing von Neumann entropy which can be approximated efficiently
by MPS.  Therefore, fix $0<\kappa<1$ and $\epsilon>0$, and set
$p_N=N^{-\epsilon(1-\kappa)}$ in (\ref{eq:magicstate}).  
As in the previous example, $p_N\rightarrow0$ implies approximability,
and
\[
S(\rho_{2N}^L)=H({p_N,1-p_N})+p_N\log[2^L] \ge L/N^{\epsilon(1-\kappa)}\ ,
\]
which implies $S(\rho_{2N}^L)\ge L^\kappa$ for $L\ge N^\epsilon$.

We now construct a state which obeys a strict area law for the
von Neumann entropy but yet cannot be approximated by MPS. Therefore,
set $M=2N^3$ and define $\ket{\chi_{M}}=\ket{\psi_{2N}}^{\otimes N^2}$
with
$\ket{\psi_{2N}}$ from (\ref{eq:magicstate}), where $p_N=1/N$.  
Then, $S(\rho_M^L)$ is at most twice the maximum
entropy of a cut through $\ket{\psi_{2N}}$, and thus 
$$
S(\rho_M^L)\le2\left(H(p_N,1-p_N)+p_N N\right)\le4\ .
$$
To prove hardness of approximation, observe that for a given $D$,
the best $D$-MPS approximation to $\ket{\psi_{2N}}^{\otimes N^2}$ 
also carries this product structure,
$\ket{\phi_D}^{\otimes N^2}$~%
\footnote{
We prove that the optimal $D$-MPS approximating
$\ket{\psi_A}\ket{\psi_B}\in\mathcal H_K\otimes\mathcal H_L$ ($K+L=N$) 
can always be chosen to carry the same product structure: 
Given a $D$-MPS $\ket{\phi_D}$ as in (\ref{eq:mps-rep}), 
write it as $\ket{\phi_D}=\sum_k\ket{\alpha_k}\ket{\beta_k}$,
with
\[
\ket{\alpha_k}=\sum A_{i_1}^{[1]}\cdots
	A_{i_K}^{[K]}e_k\ket{i_1,\dots,i_K}\ ,
\]
\[
\quad\ \ket{\beta_k}=\sum e_k^\dagger A_{i_{K+1}}^{[K+1]}\cdots
    A_{i_N}^{[N]}\ket{i_1,\dots,i_N}\ ,
\]
where $e_k$ is the $k$'th unit vector. Since the $\ket{\alpha_k}$ 
($\ket{\beta_k}$) differ
only by one boundary condition, any superposition thereof, 
 and in particular the orthonormal vectors
$\ket{\tilde\alpha_k}$, $\ket{\tilde\beta_k}$ appearing in 
the Schmidt decomposition
$\ket{\phi_D}=\sum_k\lambda_k\ket{\tilde\alpha_k}\ket{\tilde\beta_k}$,
are $D$-MPS.
  Define
$a_k:=\langle\psi_A\ket{\tilde\alpha_k}$,
$b_k:=\langle\psi_B\ket{\tilde\beta_k}$, and the factorizing $D$-MPS
\[\ket{\phi'_D}=\left(\sum_k\lambda_k
\tfrac{|a_k|}{a_k}\ket{\tilde\alpha_k}\right) \left(\tfrac{1}{|\mathcal
L|^{1/2}}\sum_{l\in\mathcal L}\tfrac{|b_l|}{b_l}
\ket{\tilde\beta_l}\right)\ ,\]
where $\mathcal L=\{l:|b_l|\ge|b_j|\,\forall
j\}$.  Then, $\ket{\phi'_D}$ is a normalized $D$-MPS, and
$|\langle\psi_A,\psi_B\ket{\phi_D}|< |\langle\psi_A,\psi_B\ket{\phi'_D}|$
unless the Schmidt rank of $\ket{\phi_D}$ is one.
}. 
From the multiplicativity of the fidelity and the relations following
Eq.~(\ref{eq:before-trig}) one infers $T(\phi^{\otimes K},\psi^{\otimes
K})\ge \sqrt{K/8}\;T(\phi,\psi)$ for $T(\phi,\psi)^2\le2/K$.
Second, from the truncation error $\epsilon_N(D)$ for
$\ket{\psi_{2N}}$, $T(\phi_D,\psi_{2N})\ge(2^N-(D-1))p_N/2^N$ for any
$D$-MPS $\ket{\phi_D}$.  Together, this shows that
$D\ge2^N(1-8T(\Phi_D,\chi_M))+1$ which is exponential in the system size
$M=2N^3$.  

It is unclear how to make this example translational invariant.
However, for the adjacent cases in Table~\ref{table:intensive}, 
those examples exist: For $S\sim\log L$, take the preceding example 
and make it
translational invariant by adding a tagging system
$\ket{10\dots0}^{\otimes N^2}$ and superposing all translations.
The resulting state is hard to approximate as the translational
invariance can be broken by local projections on the tags, and since the
reduced state $\rho_N^L$ is the translational invariant mixture of the
original, tagged reduced states, the entropy is increased by at most $\log
L$.  For the case $S_\alpha\sim\mathrm{const.}$, $\alpha>1$, the
state (\ref{eq:magicstate}) with constant $p_N$ does the job.

The last two examples are of a different type: We consider $N$ spins on an
ring and equidistantly distribute $\nu$ maximally entangled pairs between
opposite sites (i.e., $k$ and $k+N/2$), while initializing all remaining
qubits to $\ket0$. The first example, obtained for $\nu=\log N$, provides
a state with $S_\infty\sim\log L$ which is approximable. It is clearly an
MPS with $D=2^{\log N}=N$,  and for any $c>0$, $S_\infty(\rho^L_{N})\ge
\left\lfloor\tfrac LN\log N\right\rfloor\ge c\log L-1$ for $L\ge c N$.  It
can be made translational invariant by superposing all translates of the
state: On the one hand, this increases the bond dimension by at most a
factor of $N$~\cite{mps-reps}, while on the other hand, the largest
eigenvalue of a block of length $N/\log N$ is $\tfrac12$, i.e., 
the $\log L$ lower bound on the $S_\infty$ entropy remains unchanged.

The second example illustrates that for any $\kappa>0$, there is a state
with $S_0\sim N^\kappa$ which cannot be approximated by MPS. Therefore,
choose $\nu=N^\kappa$: Then, $S_0(\rho_N^L)\le N^\kappa L/N+1\le
2L^\kappa$,  while inapproximability follows from the superlogarithmic
number of maximally entangled pairs.
Translational invariance is achieved by taking the superposition of all
translations for $\kappa'<\kappa$. The spectrum of a block of length
$N/N^{\kappa'}$ is broadened to $(\tfrac12,\tfrac
{N^{\kappa'}}{2N},\dots,\tfrac{N^{\kappa'}}{2N})$: this clearly increases the
truncation error, and the entropy scaling gets a $\log$ correction
$S_0(\rho_N^L)\le2L^{\kappa'}(1+(1-\kappa')\log L)$ which is bounded by
$4L^\kappa$ for properly chosen $\kappa'$ and $L$.

Let us now prove the hardness of simulating time evolutions with MPS-based
approaches, using the results obtained (cf.\ also
\cite{guifre:mixedstate}).  To this end, take a spin chain
with all spins up, and apply a critical Ising Hamiltonian with periodic
boundary conditions.  There is good evidence~\cite{cardy:ising} that in
this case  the block entropy of any block grows linearly in time, and
indeed, a lower bound $S(\rho_N^L(t))\ge 4t/3\pi+O(\log t)$ for $t\le eL/4$
can be rigorously proven~\cite{lowerbnd-prep}.  By plugging this into
(\ref{eq:thm2:bondbound}) and setting $L=4t/e$, one finds that for an error
$\delta<2e/3\pi\approx0.58$, the required bond dimension, and thus the
effort to simulate the time evolution using MPS, grows exponentially in
time.

In this work, we have explored the relation between the scaling of block 
entropies and approximability by MPS. More refined criteria 
might be obtained by considering more involved figures of
merit. For instance, the approximability proof of~\cite{faithfully} can be
adapted to \emph{smooth R\'enyi entropies}
$S_\alpha^\epsilon(\rho)=\min\{S_\alpha
(\sigma):\|\rho-\sigma\|_\tr\le\epsilon\}$~\cite{smoothrenyi}.
Then, the existence of $\alpha<1$, $\epsilon>0$, and $c>0$ s.th.\
$S_\alpha^{1/N^{1+\epsilon}}(\rho_N^L)\le c\log N$ implies
approximability~\footnote{
This is tight: The scaling, as smooth entropies are
lower bounds on their non-smooth version, and the smoothening, as 
the inapproximability example for constant von Neumann entropy also has
constant $S_\alpha^{5/N}$ entropy.  Conversely, for every
approximable state, $S_\alpha^\epsilon(\rho_N^L)$ grows at most
logarithmically for every $\epsilon$.  This is also tight---consider, e.g.,
(\ref{eq:magicstate}) with $p_N=1/N^\kappa$.
}.  Indeed, the state
(\ref{eq:magicstate}) with $p_N=1/N^2$ has linearly growing R\'enyi
entropies, while the smooth R\'enyi entropies are constant and thus
imply approximability. 

We thank K.\ Audenaert and K.~G.\ Vollbrecht 
for helpful discussions. This work 
was supported by the EU, the DFG,  and the Elite Network of Bavaria 
project QCCC.


\begin{thebibliography}{16}
\expandafter\ifx\csname natexlab\endcsname\relax\def\natexlab#1{#1}\fi
\expandafter\ifx\csname bibnamefont\endcsname\relax
  \def\bibnamefont#1{#1}\fi
\expandafter\ifx\csname bibfnamefont\endcsname\relax
  \def\bibfnamefont#1{#1}\fi
\expandafter\ifx\csname citenamefont\endcsname\relax
  \def\citenamefont#1{#1}\fi
\expandafter\ifx\csname url\endcsname\relax
  \def\url#1{\texttt{#1}}\fi
\expandafter\ifx\csname urlprefix\endcsname\relax\def\urlprefix{URL }\fi
\providecommand{\bibinfo}[2]{#2}
\providecommand{\eprint}[2][]{\url{#2}}

\bibitem[{\citenamefont{White}(1992)}]{white:DMRG-PRL}
\bibinfo{author}{\bibfnamefont{S.~R.} \bibnamefont{White}},
  \bibinfo{journal}{Phys. Rev. Lett.} \textbf{\bibinfo{volume}{69}},
  \bibinfo{pages}{2863} (\bibinfo{year}{1992}).

\bibitem[{\citenamefont{Schollw{\"o}ck}(2005)}]{schollwoeck:rmp}
\bibinfo{author}{\bibfnamefont{U.}~\bibnamefont{Schollw{\"o}ck}},
  \bibinfo{journal}{Rev.\ Mod.\ Phys.} \textbf{\bibinfo{volume}{77}},
  \bibinfo{pages}{259} (\bibinfo{year}{2005}), \eprint{cond-mat/0409292}.

\bibitem[{\citenamefont{Vidal}(2004)}]{vidal:timeevol}
\bibinfo{author}{\bibfnamefont{G.}~\bibnamefont{Vidal}},
  \bibinfo{journal}{Phys. Rev. Lett.} \textbf{\bibinfo{volume}{93}},
  \bibinfo{pages}{040502} (\bibinfo{year}{2004}), \eprint{quant-ph/0310089}.

\bibitem[{has({\natexlab{a}})}]{hastings}
\bibinfo{howpublished}{M.\ B.\ Hastings, Phys.\ Rev.\ B \textbf{76}, 035114
  (2007), cond-mat/0701055.}

\bibitem[{\citenamefont{Verstraete and Cirac}(2006)}]{faithfully}
\bibinfo{author}{\bibfnamefont{F.}~\bibnamefont{Verstraete}} \bibnamefont{and}
  \bibinfo{author}{\bibfnamefont{J.~I.} \bibnamefont{Cirac}},
  \bibinfo{journal}{Phys. Rev. B} \textbf{\bibinfo{volume}{73}},
  \bibinfo{pages}{094423} (\bibinfo{year}{2006}), \eprint{cond-mat/0505140}.

\bibitem[{\citenamefont{Audenaert}(2006)}]{audenaert:fannes}
\bibinfo{author}{\bibfnamefont{K.~M.~R.} \bibnamefont{Audenaert}}
  J. Phys. A \textbf{40}, 8127 (2007), \eprint{quant-ph/0610146}.

\bibitem[{\citenamefont{Lloyd}(1996)}]{lloyd}
\bibinfo{author}{\bibfnamefont{S.}~\bibnamefont{Lloyd}},
  \bibinfo{journal}{Science} \textbf{\bibinfo{volume}{273}},
  \bibinfo{pages}{1073} (\bibinfo{year}{1996}).

\bibitem[{has({\natexlab{b}})}]{hastings2}
\bibinfo{howpublished}{M.\ B.\ Hastings, J. Stat. Mech. P08024 (2007),
  arXiv:0705.2024.}

\bibitem[{bqp()}]{bqp-evol}
\bibinfo{howpublished}{
  K.\ G.\ Vollbrecht and I.\ Cirac, 
    Phys. Rev. A \textbf{73}, 012324 (2006), quant-ph/0502143;
  R. Raussendorf, Phys.\ Rev.\ A \textbf{72}, 052301 (2005),
    quant-ph/0505122; 
  K.\ G.\ Vollbrecht and I.\ Cirac, Phys. Rev. Lett. \textbf{100}, 
    010501 (2008), arXiv:0704.3432;
  A.\ Kay, 
    Phys. Rev. A \textbf{76}, 030307(R) (2007),
    arXiv:0704.3142.}


\bibitem[{ksz()}]{ksz}
\bibinfo{howpublished}{A.\ Kl\"umper, A.\ Schadschneider, and J.\ Zittartz, J.
  Phys. A \textbf{24}, L955 (1991); Z. Phys. B \textbf{87}, 281 (1992).}

\bibitem[{\citenamefont{Perez-Garcia et~al.}(2007)\citenamefont{Perez-Garcia,
  Verstraete, Wolf, and Cirac}}]{mps-reps}
\bibinfo{author}{\bibfnamefont{D.}~\bibnamefont{Perez-Garcia}},
  \bibinfo{author}{\bibfnamefont{F.}~\bibnamefont{Verstraete}},
  \bibinfo{author}{\bibfnamefont{M.~M.} \bibnamefont{Wolf}}, \bibnamefont{and}
  \bibinfo{author}{\bibfnamefont{J.~I.} \bibnamefont{Cirac}},
  \bibinfo{journal}{Quant. Inf. Comput.} \textbf{\bibinfo{volume}{7}},
  \bibinfo{pages}{401} (\bibinfo{year}{2007}), \eprint{quant-ph/0608197}.

\bibitem[{\citenamefont{Bhatia}(1996)}]{bhatia}
\bibinfo{author}{\bibfnamefont{R.}~\bibnamefont{Bhatia}},
  \emph{\bibinfo{title}{Matrix Analysis}} (\bibinfo{publisher}{Springer, New
  York}, \bibinfo{year}{1996}).

\bibitem[{\citenamefont{Datta and Vidal}(2007)}]{guifre:mixedstate}
\bibinfo{author}{\bibfnamefont{A.}~\bibnamefont{Datta}} \bibnamefont{and}
  \bibinfo{author}{\bibfnamefont{G.}~\bibnamefont{Vidal}},
  \bibinfo{journal}{Phys.\ Rev.\ A} \textbf{\bibinfo{volume}{75}},
  \bibinfo{pages}{042310} (\bibinfo{year}{2007}), \eprint{quant-ph/0611157}.

\bibitem[{\citenamefont{Calabrese and Cardy}(2005)}]{cardy:ising}
\bibinfo{author}{\bibfnamefont{P.}~\bibnamefont{Calabrese}} \bibnamefont{and}
  \bibinfo{author}{\bibfnamefont{J.}~\bibnamefont{Cardy}},
  \bibinfo{journal}{J.\ Stat.\ Mech.\ P04010}  (\bibinfo{year}{2005}),
  \eprint{cond-mat/0503393}.

\bibitem[{\citenamefont{{Schuch \textit{et al.} }}()}]{lowerbnd-prep}
    N.\ Schuch, M.\ M.\ Wolf, K.\ G.\ Vollbrecht, and J.\ I.\ Cirac
    (2008), arXiv:0801.2078.

\bibitem[{\citenamefont{Renner}(2005)}]{smoothrenyi}
\bibinfo{author}{\bibfnamefont{R.}~\bibnamefont{Renner}}
  (\bibinfo{year}{2005}), \eprint{quant-ph/0512258}.

\end{thebibliography}
\end{document}